\documentclass[aps,twocolumn,prc,superscriptaddress,showpacs]{revtex4}

\usepackage{graphicx}
\usepackage{dcolumn}
\usepackage{bm}

\usepackage{amsmath}    
\usepackage{graphicx}   

\begin{document}

\title{Fission barriers in actinides in covariant density functional theory:
the role of triaxiality.}

\author{H. Abusara}
\affiliation{Department of Physics and Astronomy, Mississippi State
University, MS 39762}
\author{A. V. Afanasjev}
\affiliation{Department of Physics and Astronomy, Mississippi State
University, MS 39762}
\author{P.\ Ring}
\affiliation{Fakult\"at f\"ur Physik, Technische Universit\"at M\"unchen,
 D-85748 Garching, Germany}

\date{\today}

\begin{abstract}
  Relativistic mean field theory allowing for triaxial deformations is
applied for a systematic study of fission barriers in the actinide
region. Different pairing schemes are studied in details and it is shown
that covariant density functional theory is able to describe fission
barriers on a level of accuracy comparable with non-relativistic
calculations, even with the best phenomenological macroscopic+microscopic
approaches. Triaxiality in the region of the first saddle plays a crucial
role in achieving that.
\end{abstract}

\maketitle

\section{Introduction}

  A study of the (static) inner fission barrier heights $B_{f}^{st}$
of even-even nuclei is motivated by the importance of this quantity for
several physical phenomena. Many heavy nuclei decay by spontaneous fission,
and the size of the fission barrier is a measure for the stability of a
nucleus reflected in the spontaneous fission lifetimes of these
nuclei~\cite{SP.07}. The probability for the formation of a
superheavy nucleus in a heavy-ion-fusion reaction is also directly
connected to the height of its fission barrier~\cite{IOZ.02}. The
height $B_{f}^{st}$ is a decisive quantity in the competition between
neutron evaporation and fission of a compound nucleus in the process
of its cooling. The large sensitivity of the cross section $\sigma$
for the synthesis of the fissioning nuclei on the barrier height
$B_{f}^{st}$ stresses a need for accurate calculations of this value.
For example, a change of $B_{f}^{st}$ by 1 MeV changes the calculated
survival probability of a synthesized nucleus by about one order of
magnitude or even more~\cite{IOZ.02}. The population and survival of
hyperdeformed states at high spin also depends on the fission
barriers~\cite{DPS.04,AA.08}. In addition, the $r-$process of stellar
nucleosynthesis depends (among other quantities such as masses and
$\beta$-decay rates) on the fission barriers of very neutron-rich
nuclei~\cite{AT.99,MPR.01}.

  During the last decade the role of triaxiality in the region of the
saddle point of fission barriers has been recognized and tested in
many theoretical frameworks. It was found that the height of the
barrier is reduced when triaxial shapes are
allowed~\cite{SBDN.09,WERP.02}. However, this lowering strongly
depends on the proton and neutron numbers and on the model employed.
The investigations of inner fission barriers with triaxiality
included are available within the frameworks of the
microscopic+macroscopic
method~\cite{MSI.04,SK.06,DPB.07,MSI.09,DNPB.09}, the extended
Thomas-Fermi plus Strutinsky integral~\cite{DPT.00}, and
non-relativistic energy density functionals based on
Skyrme~\cite{BRRMG.98,BQS.04,SDN.06,SBDN.09} and
Gogny~\cite{WERP.02,War.09,RSRG.10} forces.

  Covariant density functional theory (CDFT)~\cite{VALR.05} is an
approach alternative to the above mentioned non-relativistic methods.
Built on Lorentz covariance and the Dirac equation, it provides a
natural incorporation of spin degrees of freedom~\cite{Rei.89,Rin.96}
and an accurate description of spin-orbit splittings~\cite{Rin.96}
(see also Fig. 2 in Ref.\ \cite{BRRMG.99}), which has an essential
influence on the underlying shell structure. Lorentz covariance of
the CDFT equations leads to the fact that time-odd mean fields of
this theory are determined as spatial components of Lorentz vectors
and therefore coupled with the same constants as the time-like
components~\cite{AA.10} which are fitted to ground state properties
of finite nuclei. In addition, pseudo-spin symmetry finds a natural
explanation in the relativistic framework~\cite{Gin.97}. Over the
years a large variety of nuclear phenomena have been successfully
described within the CDFT (see Ref.\ \cite{VALR.05}
and references therein).

  However, the progress in the study of the fission barriers within
CDFT has been slower than in its non-relativistic counterparts. Inner
fission barriers in several nuclei have been calculated in the
axially symmetric relativistic mean field (RMF) + BCS approach in
Refs.\ \cite{BMR.94,RMRG.95,RTC.02,ZZZM.03,LGM.06}. However, these
investigations employ the constant gap approximation in the BCS part.
Our recent study of pairing schemes used for the calculations of
fission barriers clearly shows that this approximation leads to
unphysical results for the fission barriers~\cite{KALR.10}. Thus, the
results of these works have to be treated with a caution. Fission
barriers have also been studied in axially symmetric RMF calculations
within the BCS approximation using an effective density-dependent
zero-range force in the pairing channel; this force represents a much
more realistic approximation for pairing \cite{KALR.10}. Recently
also relativistic Hartree-Bogoliubov (RHB) calculations with the
Gogny force D1S and with $\delta$-forces in the pairing channel have
been carried out~\cite{KALR.10} for a study of  fission
barriers with axial symmetry.

  Unfortunately, axially symmetric calculations cannot be directly
compared with experimental data since, as has been shown in
non-relativistic calculations~\cite{SBDN.09,WERP.02}, the lowering of
fission barriers due to triaxiality is significant and can reach 3-4
MeV in some nuclei. At present, no systematic studies of the effects
of triaxial degrees of freedom on the height of inner fission
barriers are available in the covariant density functional theory;
this degree of freedom has only been studied in specific nuclei such
as $^{264}$Hs \cite{BRRMG.98} and  $^{240}$Pu \cite{BHR.03} within the
RMF+BCS approach as well as $^{240}$Pu \cite{LNV.10} within the
RHB approach. Thus, the main goal
of the current manuscript is to perform a systematic investigation of
the inner fission barriers within the triaxial RMF+BCS approach, and
for the first time to confront these important experimental
quantities with CDFT in a systematic way.

The manuscript is organized as follows. The triaxial RMF+BCS theory
and its details related to the calculations of fission barriers are
discussed in Sec.\ \ref{Theor-sect}. Sec.\ \ref{Trun-ph} is devoted
to the analysis of the effects of the truncation of the basis in the
particle-hole channel of the model. Truncation effects in the pairing
channel are considered in Sec.\ \ref{Trun-pai}. The results of the
calculation of the fission barriers, the role of triaxiality and the
comparison with experiment are discussed in Sec.\ \ref{Sys-sec}.
Finally, in Sec.\ \ref{Sec-DD}, we report on calculations with others
relativistic parameter sets based on density dependent coupling
constants and in Sec.\ \ref{Sec-final} summarize the results of our
work.

\section{Theoretical framework and the details of numerical calculations}
\label{Theor-sect}

  The starting point of Covariant Density Functional Theory (CDFT) is a
standard Lagrangian density~\cite{GRT.90}
\begin{align}
\mathcal{L}  &  =\bar{\psi}\left(
\gamma(i\partial-g_{\omega}\omega-g_{\rho
}\vec{\rho}\vec{\tau}-eA)-m-g_{\sigma}\sigma\right)  \psi\nonumber\\
&  +\frac{1}{2}(\partial\sigma)^{2}-\frac{1}{2}m_{\sigma}^{2}\sigma^{2}%
-\frac{1}{4}\Omega_{\mu\nu}\Omega^{\mu\nu}+\frac{1}{2}m_{\omega}^{2}\omega
^{2}\label{lagrangian}\\
&  -\frac{1}{4}{\vec{R}}_{\mu\nu}{\vec{R}}^{\mu\nu}+\frac{1}{2}m_{\rho}%
^{2}\vec{\rho}^{\,2}-\frac{1}{4}F_{\mu\nu}F^{\mu\nu}\nonumber
\end{align}
which contains nucleons described by the Dirac spinors $\psi$ with
the mass $m$ and several effective mesons characterized by the
quantum numbers of spin, parity, and isospin. They create effective
fields in a Dirac equation, which corresponds to the Kohn-Sham
equation~\cite{KS.65} in the non-relativistic case.

  The Lagrangian (\ref{lagrangian}) contains as parameters the meson
masses $m_{\sigma}$, $m_{\omega}$, and $m_{\rho}$ and the coupling
constants $g_{\sigma}$, $g_{\omega}$, and $g_{\rho}$. $e$ is the
charge of the protons and it vanishes for neutrons. This model has
first been introduced by Walecka~\cite{Wal.74,SW.86}. It has turned
out that surface properties of finite nuclei cannot be described
properly by this model. Therefore, Boguta and Bodmer~\cite{BB.77}
introduced a density dependence via a non-linear meson coupling
replacing the term $\frac{1}{2}m_{\sigma}^{2}\sigma^{2}$ in Eq.
(\ref{lagrangian}) by
\begin{equation}
U(\sigma)~=~\frac{1}{2}m_{\sigma}^{2}\sigma^{2}+\frac{1}{3}g_{2}\sigma
^{3}+\frac{1}{4}g_{3}\sigma^{4}.
\end{equation}
If not specified otherwise, the calculations are performed with the
NL3* parameterization of the RMF Lagrangian \cite{NL3*} shown in
Table \ref{tab1}.  Apart from the fixed values for the
masses $m=939$ MeV, $m_\omega=782.6$ MeV and $m_\rho=763$ MeV it
contains six phenomenological parameters $m_\sigma$, $g_\sigma$,
$g_\omega$, $g_\rho,g_2$, and $g_3$, which have been recently
adjusted to the experimental data in finite nuclei\ \cite{NL3*}
eliminating a few deficiencies of the well known older parameter set
NL3\ \cite{NL3}.

\begin{table}[ptb]
\caption{{\protect\small Parameters of the effective interaction NL3*
in the RMF Lagrangian}}%
\label{tab1}
\begin{center}%
\begin{tabular}
[c]{rlrl}\hline\hline & Parameters of NL3* &  & \\\hline
$m$ & = 939 (MeV) &  & \\
$m_{\sigma}$ & = 502.5742 (MeV) & $g_{\sigma}$ & = 10.0944\\
$m_{\omega}$ & = 782.600 (MeV) & $g_{\omega}$ & = 12.8065\\
$m_{\rho}$ & = 763.000 (MeV) & $g_{\rho}$ & = 4.5748\\
$g_{2}$ & = -10.8093 (fm$^{-1}$) &  & \\
$g_{3}$ & = -30.1486 &  & \\ \hline\hline
\end{tabular}
\end{center}
\end{table}

  In the current investigation, the triaxial RMF+BCS model is
used~\cite{KR.88}. The RMF-equations are solved and at each step of
the iteration the BCS occupation probabilities $v_{k}^{2}$ are
determined. These quantities are used in the calculation of
densities, energies and new fields for the next step of the
iteration. We use monopole pairing force with the strength parameters
$G_{\tau}$ for neutrons ($\tau=n$) and protons ($\tau=p$); this
method is based on the residual interaction of the seniority model
\cite{RS.80}.

 We start with a pairing strength parameters $G$ and solve in each step of
the iteration the gap equation~\cite{RS.80}
\begin{equation}
\frac{1}{G}=%
\sum\limits_{k>0}\frac{1}{2E_{k}}%
\label{gap-equation}%
\end{equation}
with $E_{k}=\sqrt{(\varepsilon_{k}-\lambda)^{2}+\Delta^{2}}$,  where
$\varepsilon_{k}$ are the eigenvalues of the Dirac equation and
the chemical potential $\lambda$ is determined by the average
particle number. Then the occupation probabilities
\begin{equation}
v_{k}^2=\frac{1}{2}\left(  1-\frac{\varepsilon_{k}-\lambda}{E_{k}%
}\right),  \label{v2}%
\end{equation}
and the gap parameters
\begin{equation}
\Delta=G\sum\limits_{k>0}u_{k}v_{k}%
\label{delta-BCS}%
\end{equation}
are determined in a self-consistent way. The pairing energy is defined
as
\begin{equation}
E_{\rm pair}=-\Delta%
\sum\limits_{k>0}
u_{k}v_{k}, %
\label{Epair}%
\end{equation}
The sum over $k$ in Eqs.~(\ref{gap-equation}), (\ref{delta-BCS}) and (\ref{Epair})
run over all states in the pairing window $E_k<E_{\rm cutoff}$.%

In Ref.\ \cite{MN.92} empirical pairing gap parameters
\begin{eqnarray}
\Delta_{n}^{emp} = \frac{4.8}{N^{1/3}}\,\,\,\, {\rm MeV}, \qquad
\Delta_{p}^{emp} = \frac{4.8}{Z^{1/3}}\,\,\,\, {\rm MeV}
\end{eqnarray}
have been determined by the systematic fit to experimental data on
neutron and proton gaps in the normal deformed minimum.

  These empirical gap parameters form the basis for the definition of
the strength parameters $G_{\tau}$ in the current manuscript. Two
procedures have been used: a) In the analysis of different truncation
(Sec.\ \ref{Trun-ph}) and pairing schemes (Sec.\ \ref{Trun-pai}), the
values $G_n(Z,N)$ and $G_p(Z,N)$ are defined for each nucleus with
neutron and proton number $N$ and $Z$ under study from the
requirement that, in the normal deformed minimum, the calculated
pairing gaps coincide with the empirical values. b) In the systematic
calculations of potential energy surfaces and fission barriers in
actinides the same procedure is used first for all even-even nuclei in
the $Z=90-100$ and $N-Z=42-66$ ranges resulting  in a set of the
strengths $G_n(Z,N)$ and $G_p(Z,N)$. Then, the following expressions
\cite{DMS.80}
\begin{eqnarray}
A \cdot G_n = G_1^n - G_2^n \frac{N-Z}{A}\,\,\,\,{\rm MeV} \\
A \cdot G_p = G_1^p + G_2^p \frac{N-Z}{A}\,\,\,\,{\rm MeV}
\label{G-strength}
\end{eqnarray}
are used in the calculations. The parameters $G_1^n$, $G_2^n$, $G_1^p$
and $G_2^p$ are defined by the least square fit to the set of the
$G_n(Z,N)$ and $G_p(Z,N)$. Their values depend on the parameter set of
the Lagrangian and they are given in Table \ref{tab2}. In this way
we have strength parameters for the effective pairing interaction
depending in a smooth way on the neutron and proton numbers and,
because of the changing level density, the gap parameters derived
from those values show fluctuations as a function of the particle
numbers.

\begin{table}[t]
\caption{The $G_1^n$, $G_2^n$, $G_1^p$ and $G_2^p$ parameters [in
MeV] for different parameterizations of the RMF Lagrangian and cutoff
energy $E_{\rm cutoff}=120$ MeV.} \label{tab2}
\renewcommand{\tabcolsep}{0.5pc}
\renewcommand{\arraystretch}{1.4}
\begin{tabular} {ccccc}\hline
Force  & $G_1^n$ &  $G_2^n$  &  $G_1^p$ & $G_2^p$ \\ \hline
NL3*   &  9.1    &   6.4     &   8.1    & 10.0   \\
DD-PC1 &  9.2    &   5.4     &   8.0    & 11.4   \\
DD-ME2 &  9.2    &   5.8     &   8.1    & 11.2   \\ \hline
\end{tabular}
\\[2pt]\end{table}

  The calculations are performed imposing constraints on the axial and
triaxial mass quadrupole moments. The method of quadratic constraints
uses a variation of the function
\begin{eqnarray}
\left<H\right> + \sum_{\mu=0,2} C_{2\mu}
(\langle\hat{Q}_{2\mu}\rangle-q_{2\mu})^2
\end{eqnarray}
where $\left<H\right>$ is the total energy, and
$\langle\hat{Q}_{2\mu}\rangle$ denotes the expectation values of the
mass quadrupole operators
\begin{eqnarray}
\hat{Q}_{20}&=&2z^2-x^2-y^2 \\
\hat{Q}_{22}&=&x^2-y^2
\end{eqnarray}
  In these equations, $q_{2\mu}$ is the constrained value of the multipole moment,
and $C_{2\mu}$ the corresponding stiffness constants \cite{RS.80}.

  Correlations beyond mean field can influence the calculated values of
fission barrier height and the excitation energies of the
superdeformed minima associated with the fission isomer
\cite{BHB.04}. The inclusion of rotational correlations can be
performed by a symmetry restoration (angular momentum projection) and
that of vibrations by a mixing of mean field states corresponding to
different shapes by the method of generator coordinates (GCM). So far
such an investigation has been performed only for $^{240}$Pu within
the generator coordinate method based on Skyrme DFT under the
restriction to axially symmetric shapes \cite{BHB.04}. It was found
that compared to the ground state, angular momentum projection lowers
the (axial) inner barrier by about 0.6 MeV and the fission isomer by
about 1 MeV. In addition, it was found in Ref.\ \cite{BHB.04} that the
schematic rotational correction based on the Belyaev moment of
inertia~\cite{Bel.59}
frequently used in the literature gives a reduction of the fission
barrier height which is appreciably larger than the one due to
angular momentum projection. Based on these results, no rotational
corrections are taken into account in our calculations. A similar
approach has been used in the very successful calculations of
Ref.\ \cite{MSI.09}.

  In the current investigation we do not consider the outer fission
barriers. However, this restriction has its own merits. The inner
barriers are generally better measured than the outer ones, and they
are certainly more important for the $r-$ process, since they
determine thresholds. Furthermore, spontaneous fission lifetimes tend
to be dominated by the inner barrier, even if occasionally an outer
barrier can have a crucial effect if it is wide enough. The
consideration of only inner fission barriers allows us to restrict
our calculations to reflection symmetric shapes, because the
odd-multipole deformations (octupole, etc.) do not play a role in the
inner fission barrier of the actinides and superheavy nuclei
\cite{RMRG.95,SDN.06,SK.06,CDHMN.96,SGP.05}. However, the
analysis of the symmetric fission pathway, which is the lowest in
energy in heavy actinide nuclei such as some Fm isotopes (see
Refs.\ \cite{BRRMG.98,SBDN.09}), is also possible in the current
framework.

\section{Truncation effects in  the particle-hole channel}
\label{Trun-ph}

  The RMF+BCS equations are solved in the basis of an anisotropic
three-dimensional harmonic oscillator in Cartesian coordinates
characterized by the deformation parameters $\beta_0$ and $\gamma_0$ and
the oscillator frequency $\hbar \omega_{0}=41A^{-1/3}$ MeV (see
Refs.\ \cite{KR.88,AKR.96} for details). The deformation
parameters of the oscillator basis $\beta_0$ and $\gamma_0$ are
selected to be close to expected values $\beta_2$ and $\gamma$ of
constrained solution; this improves the convergence and minimizes the
computational time. The truncation of the basis is performed in such
a way that all states belonging to the shells up to $N_{F}$ fermionic
shells and $N_{B}$ bosonic shells are taken into account. The
computational time increases considerably with the increase of
$N_{F}$ but it is much less dependent on $N_{B}$. Thus, special
attention has been paid to the selection of $N_{F}$ of the basis to
be chosen for a systematic study of fission barriers for the nuclei
of interest, which provides at the same time a reasonable numerical
accuracy in the predictions of the physical observables.

  The selection of the truncation scheme was guided by the detailed
analysis of the convergence performed in axially symmetric RMF+BCS
and RHB calculations of Ref.\ \cite{KALR.10}. In this reference,
extensive tests of numerical convergence have been performed in the
spherical, normal-deformed and superdeformed ($\beta_{2}\sim0.7-1.0$)
minima in the RMF calculations without pairing on the example of the
nuclei $^{238}$U and $^{304}120$ with $Z=120$ and $N=184$. Contrary
to the previous studies of the convergence in the RMF framework which were
based on the comparison of the $N_{F}$ and $N_{F}+2$ results, the
\textquotedblleft exact\textquotedblright\ solution (extending the
calculations up to $N_{F}=36,N_{B}=36$) has been defined. Then it was
shown that the binding energies and inner fission barriers for
$N_{F}=20$ and $N_{B}=20$ were described with an accuracy of
approximately 200 keV and  100 keV, respectively, as compared with
the exact solution. Therefore, the systematic calculations in the
present manuscript have been carried out with $N_F=20$ and $N_B=20$.
This selection of the basis is in line with our previous
convergence tests in different mass regions (see, for example, Ref.\
\cite{AA.08}) which clearly show that at large deformations full
convergence of the binding energies is reached at larger values of
$N_F$ than at lower deformations. In addition, they show that larger
sizes of the basis (larger $N_F$ and $N_B$ values) are needed for the
nuclei with larger proton $Z$ and neutron $N$ numbers (see Refs.\
\cite{ARR.99,AKR.96,AA.10}).

  Of course, as long as the same number of fermionic shells $N_F$
and the same deformation of the basis $\beta_0$ is used, calculations
with the axial code should give identical results to those obtained
with the triaxial code at $\gamma=0^{\circ}$. In fact for nuclei under 
study we find agreement with an accuracy of approximately 50 keV 
throughout the deformation range of interest, which is caused by small 
differences in the mesh points of the Gaussian integrations for the 
matrix elements. As a result for axially symmetric shapes, the fission 
barrier heights which depend on the relative energies of the saddle point 
and normal deformed minimum differ by less than 50 keV in these two 
calculations.

\begin{figure}[ht]
\includegraphics[width=8.0cm]{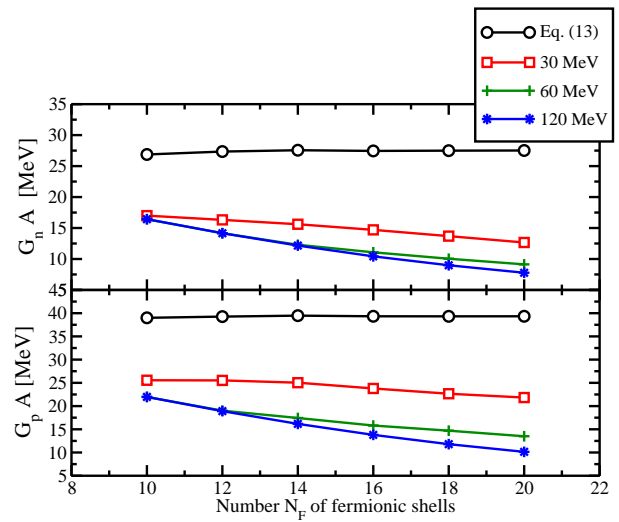}
\caption{(Color online) Neutron and proton pairing strengths as a function
of number $N_F$ of fermionic shells for different pairing schemes. Pairing
schemes are indicated either by cut-off energy $E_{\rm cutoff}$ or by
Eq.\ (\ref{eq:cutoff}).}
\label{Gpair-Nf}
\end{figure}

\begin{figure}[ht]
\includegraphics[width=8.0cm]{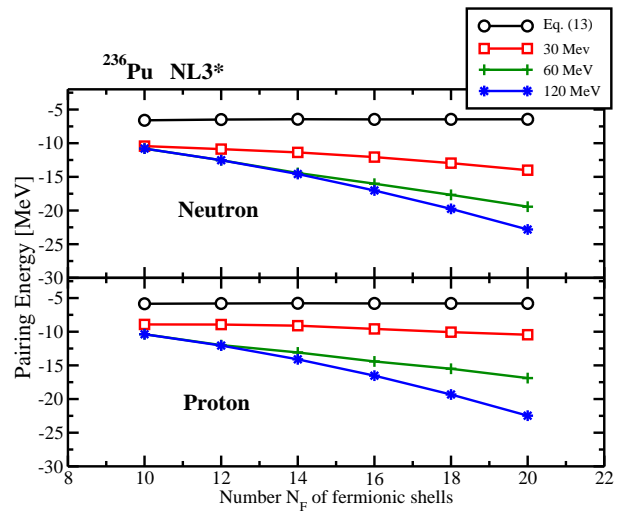}
\caption{(Color online) Neutron and proton pairing energies in
the normal deformed minimum as a function of number $N_F$ of
fermionic shells for different pairing schemes.} \label{Epair-Nf}
\end{figure}

  Extensive convergence tests in axially symmetric Skyrme-Hartree-Fock
calculations also show that a similar size of the basis is needed
(see Sec.\ IIB in Ref.\ \cite{SGP.05} for more details). The
comparison of these convergence tests suggests that there is no big
difference in the convergence of total energies as a function of the
size of basis in the relativistic and non-relativistic approaches.

\begin{figure}[ht]
\includegraphics[width=8.0cm]{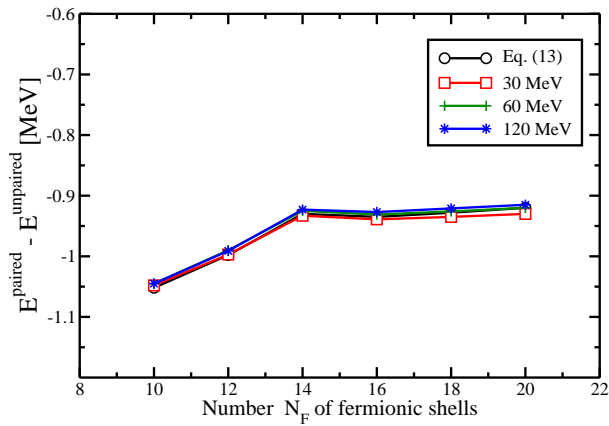}
\caption{(Color online) The dependence of additional binding due to
pairing on the number $N_F$ of major fermionic shells for different
pairing schemes.}
\label{E-pair-nopair}
\end{figure}

  Because the fermionic basis contains large and small components of
the Dirac spinor, the diagonalization of the Dirac equation is by a
factor of approximately 8 more time consuming than the corresponding
Schr\"{o}dinger equation in the non-relativistic case. As a consequence,
triaxial RMF+BCS calculations are more computationally demanding than
the ones performed in the triaxial Skyrme EDF with BCS approximation
of Ref.\ \cite{SBDN.09}. This is also a reason why we treat the pairing
channel in the present triaxial RMF calculations in the BCS
approximation despite the fact that triaxial cranked Relativistic
Hartree+Bogoliubov approach has been developed in the end of
nineties~\cite{AKR.99,ARK.00,AKRRE.00} and successfully applied to the
description of rotational structures in the pairing regime in
different mass regions \cite{AKR.99,ARK.00,AKF.03,AF.05b,AKRRE.00}. The
RMF+BCS calculations are less time-consuming than the RHB
calculations. In addition, as follows from our experience of the
calculations in axially deformed RMF+BCS and RHB codes
\cite{KALR.10}, the RMF+BCS calculations are more stable (especially,
in the saddle point region) than the RHB calculations.

\section{Truncation effects in the pairing channel}
\label{Trun-pai}

 It is rather customary to analyze the dependence of total binding
energies (or other physical observables) on the truncation of basis
(see Sec.\ \ref{Trun-ph}). However, we were not able to find any
detailed investigation where the impact of the size of the oscillator
basis on the parameters of pairing in the BCS framework has been
discussed in detail. Thus, we studied the dependence of the strength
of the pairing interaction on the number $N_F$ of fermionic shells
under the condition that the proton and neutron pairing gaps in
the normal deformed minimum are fixed for all values of $N_F$ and
cut-off energies $E_{\rm cutoff}$.

\begin{figure}[h]
\includegraphics[width=8.0cm]{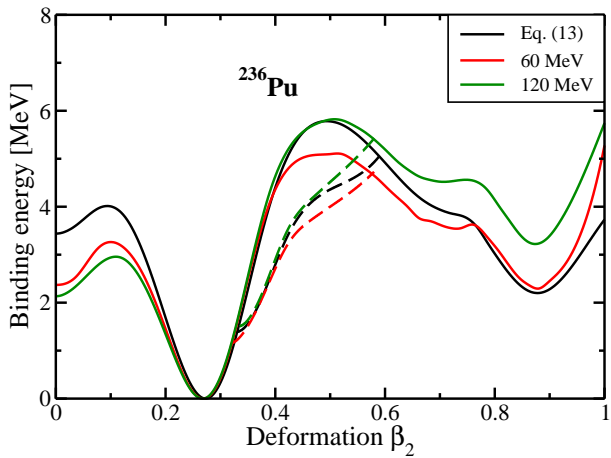}
\caption{(Color online) The dependence of fission barrier in $^{236}$Pu on
the pairing scheme. Solid and dashed lines are used for axially symmetric and
triaxial solutions, respectively.}
\label{E-pairsch-236Pu}
\end{figure}

\begin{figure}[h]
\includegraphics[width=8.0cm]{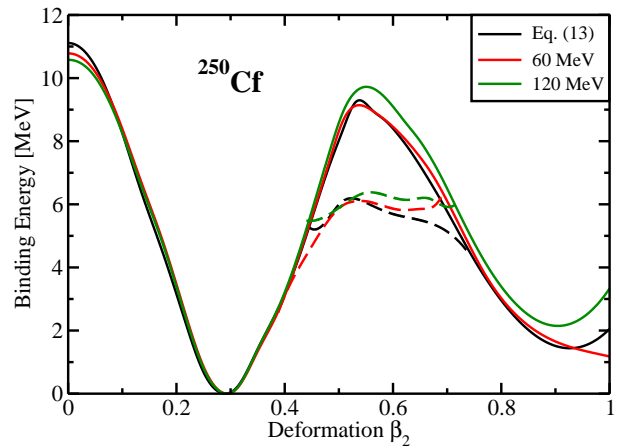}
\caption{(Color online) The same as in Fig.\ \ref{E-pairsch-236Pu} but
for $^{250}$Cf.}
\label{E-pairsch-250Cf}
\end{figure}

 RMF+BCS calculations in the normal-deformed minimum have been
performed with several values of the cut-off energy $E_{\rm cutoff}$
in Eq. (\ref{delta-BCS}), namely $E_{\rm cutoff}=30$, 60, and  120
MeV. In addition, the prescription of Ref.\ \cite{BRRM.00b}
(indicated as ``Eq. (\ref{eq:cutoff})'' in the figures) has been used.
This prescription introduces smooth energy-dependent cut-off weights
\cite{KBF.90}
\begin{equation}
\label{eq:cutoff} f_k = \frac{1}{1 + \exp [(\epsilon_k - \lambda_\tau
- \Delta E_\tau) / \mu_\tau] }
\end{equation}
for the evaluation of the local pair density. In this equation,
$\epsilon_k$ are the eigenvalues of the Dirac equation and the
chemical potentials $\lambda_\tau$ of the proton ($\tau=p$) or
neutron ($\tau=n$) subsystems are determined by the particle numbers
$N_\tau$. The cut-off parameters $\Delta E_\tau$ and $\mu_\tau =
\Delta E_\tau/10$ are chosen self-adjusting to the actual level
density in the vicinity of the Fermi energy. $\Delta E_\tau$ is fixed
from the condition that the sum of the cut-off weights includes
approximately one additional shell of single-particle states above
the Fermi surface
\begin{equation}
\sum_{k \in \Omega_\tau} f_k = N_\tau + 1.65 \, N_\tau^{2/3} \quad .
\label{eq-sum}
\end{equation}
In Eq.\ (\ref{eq-sum}), $\Omega_\tau$ denotes the single-particle
space used in the calculations.

\begin{figure}
\includegraphics[width=8.0cm]{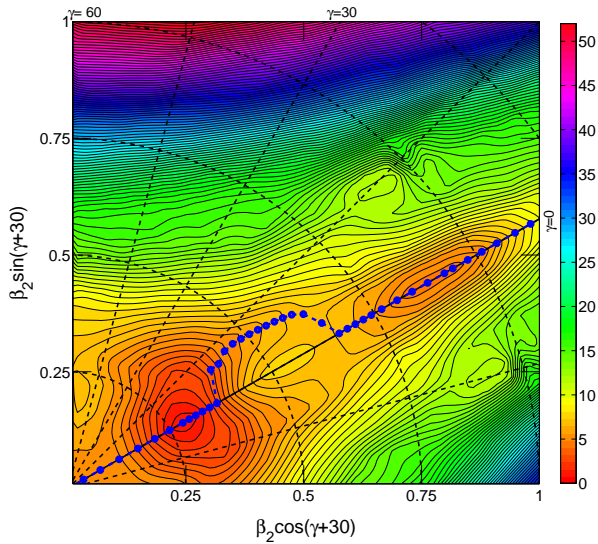}
\caption{(Color online) Potential energy surface in $^{240}$Pu. The
energy difference between two neighboring equipotential lines is
equal to 0.5 MeV. The blue dashed line with solid circles shows the
lowest in energy solution as a function of $\beta_2$. Further details
are given in the text.}
\label{pes-2D}
\end{figure}

\begin{figure*}
\includegraphics[angle=90,width=16.0cm]{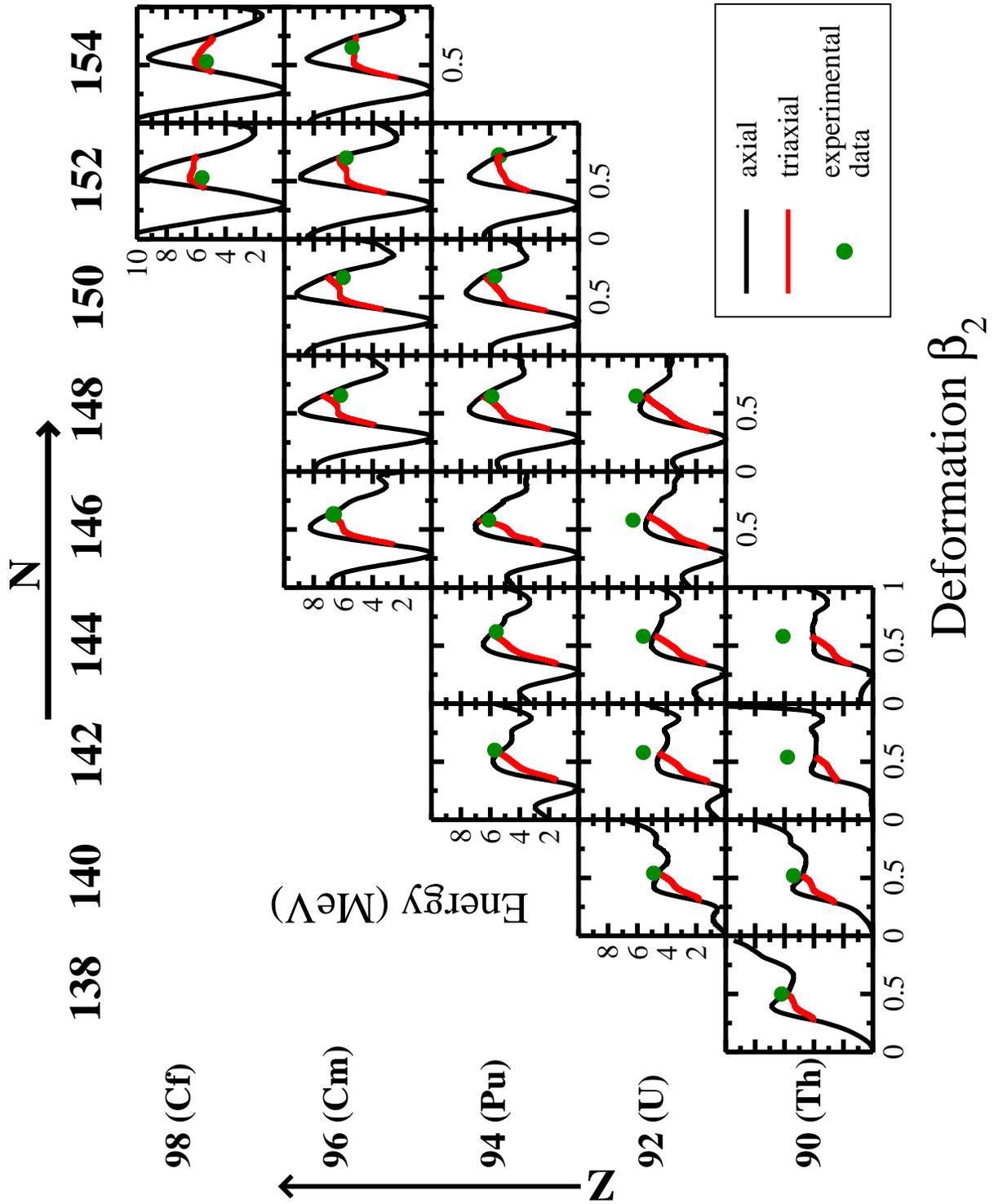}
\caption{(Color online) Deformation energy curves of even-even
actinide nuclei obtained in the RMF+BCS calculations with the NL3*
parameterization. Black solid lines display the deformation energy
curves for the axially symmetric solution, while red solid lines show
the deformation energy curves along the triaxial part of the fission
path. Solid green circles show the experimental values of the height
of inner fission barrier. Experimental data are taken from Table IV
in Ref.~\cite{SGP.05}. A typical uncertainty in the experimental
values, as suggested by the differences among various compilations,
is of the order of $\pm0.5$ MeV~\cite{SGP.05}.} \label{pot}
\end{figure*}

  Figures \ref{Gpair-Nf}, \ref{Epair-Nf} and \ref{E-pair-nopair} summarize
the results of this study for the normal-deformed ground state in $^{236}$Pu.
One can see in Fig.\ \ref{Gpair-Nf} that the strengths of the pairing interaction
depends not only on the cut-off energy $E_{\rm cutoff}$ but also on the number
$N_F$ of fermionic shells employed in the calculations. This dependence is very
weak for the prescription of Ref.\ \cite{BRRM.00b} because here the effective
pairing window is quite small being around 7 MeV. On the other hand, the dependence
of the pairing strength on $N_F$ increases with the increase of $E_{\rm cutoff}$.
This can be understood in the following way: an increase of $N_F$ brings more
single-particle states into the pairing window thus effectively
requiring the decrease of pairing strength in order to keep the
pairing gap fixed. This effect becomes more pronounced for larger
pairing windows, which explains the steeper decrease of the pairing
strength as a function of $N_F$ with increasing $E_{\rm cutoff}$.

The dependence of proton and neutron pairing energies $E_{\rm
pair}^p$ and $E_{\rm pair}^n$ on the cut-off energy $E_{\rm cutoff}$
and on the number $N_F$ of fermionic shells employed in the
calculations is shown in Fig.\ \ref{Epair-Nf}. These energies depend
only weakly on $N_F$ in the case of prescription of Ref.\
\cite{BRRM.00b} because of the small effective pairing window.
However, similar to the pairing strengths the dependence of pairing
energies on $N_F$ increases significantly with increasing $E_{\rm
cutoff}$. The origin of this feature is the same as in the case of
the pairing strengths; it is discussed above.

\begin{figure*}[ht]
\includegraphics[width=16.0cm]{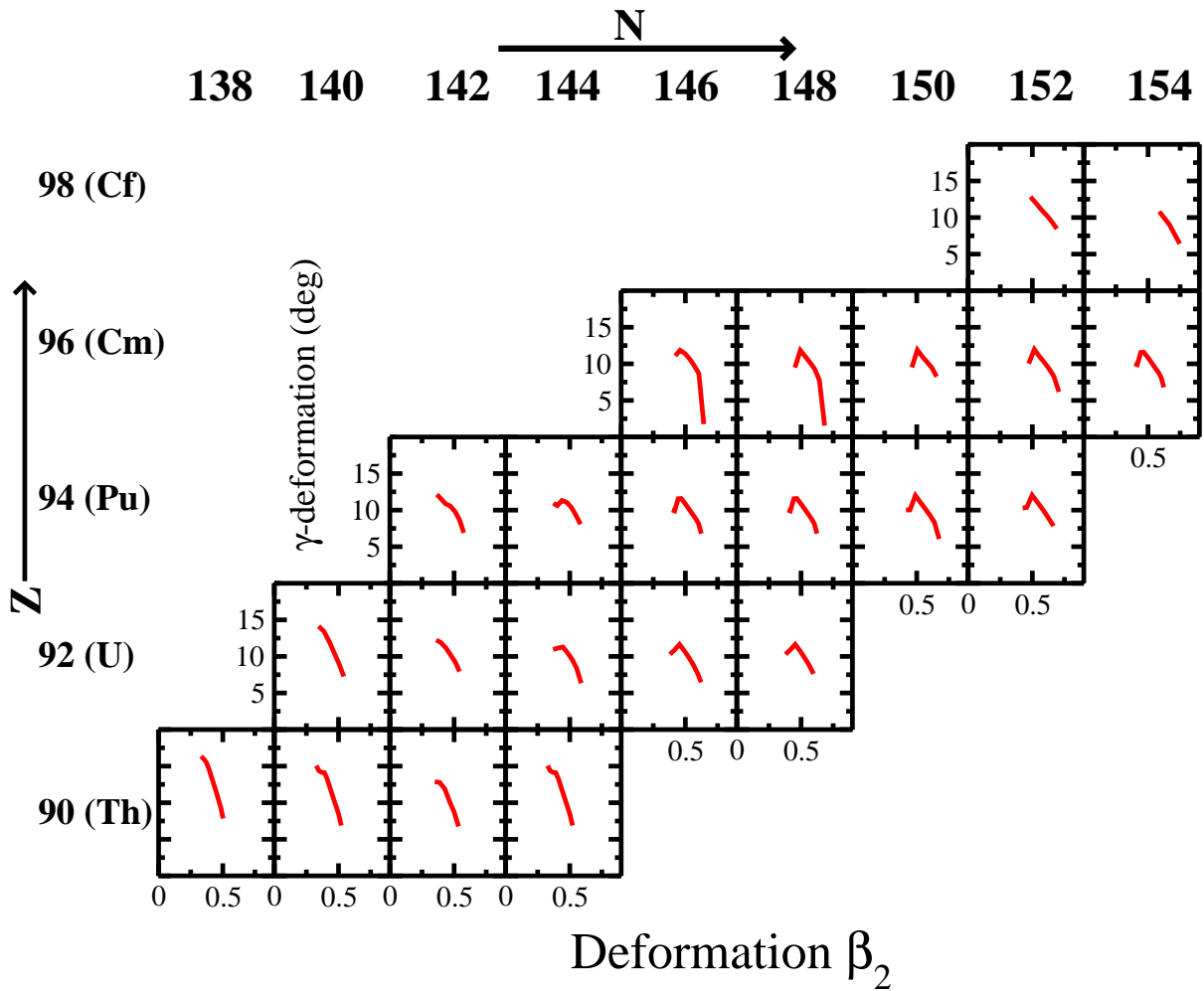}
\caption{(Color online) The $\gamma$-deformations for the results of
calculations shown by red lines in Fig.\ \ref{pot}.}
\label{pot-gamma}
\end{figure*}

Note, however, that the dramatic changes in the pairing energy cannot
be seen directly in the change of the energy, because they are
compensated to some extent by the fact that larger pairing seen in
pairing energies causes a wider distribution of the occupation
probabilities $v_k^2$ around the Fermi surface. Therefore, we study
in Fig.\ \ref{E-pair-nopair} the energy difference between the
binding energies $E^{\rm pair}-E^{\rm unpair}$ obtained in two
self-consistent calculations with and without pairing. It turns out
that for $N_F\geq 14$ this difference, that reflects the real
physical impact of pairing, is smaller than 1 MeV and it does neither
depend on the cut-off energy $E_{\rm cutoff}$ nor on the value of
$N_F$. Somewhat different values of $E^{\rm pair}-E^{\rm unpair}$ at
lower $N_F$ values are due to the fact that at these values of $N_F$
the effects of the truncation of basis in the particle-hole channel
have not been eliminated (see Sect.\ \ref{Trun-ph} for detail).

In Figs.\ \ref{E-pairsch-236Pu} and \ref{E-pairsch-250Cf} we compare
the deformation energy curves
for three different pairing schemes for the nuclei $^{236}$Pu and
$^{250}$Cf. The deformation energy curve for the axially symmetric
solution is obtained as the $\gamma=0^{\circ}$ cross-section of the
potential energy surface. The deformation energy curve for the
triaxial solution is obtained by the minimization of potential energy
surface along the $\beta_2$-direction. We show the deformation energy
curve for the triaxial solution only in the range of $\beta_2$ values
where it is lower in energy than the deformation energy curve of the
axially symmetric solution. Note that the potential energy surfaces
are normalized to zero at the normal-deformed minimum. As
discussed in Ref.~\cite{KALR.10} we can see that these different
schemes predict somewhat different fission barriers. In the
systematic calculations presented in the following sections, we use
the cut-off energy $E_{\rm cutoff}=120$ MeV. The selection of this
value is based on the results of Ref.\ \cite{KALR.10}, where it was
shown that the difference in the height of fission barriers obtained
in the RHB calculations with finite range D1S Gogny force and
zero-range $\delta$-force is minimal when the cut-off energy $E_{\rm
cutoff}=120$ MeV is used (see Fig. 6 in Ref.\ \cite{KALR.10}).

\section{Systematic analysis of the inner fission barriers}
\label{Sys-sec}

  In this section we carry out a systematic investigation of fission
barriers of even-even nuclei in the actinide region based on the
parameter set NL3* and the pairing strength parameters given in
Table~\ref{tab2}. In Fig.~\ref{pes-2D} we show as an example the
potential energy surface of the nucleus $^{240}$Pu in the
$\beta$-$\gamma$ plane. For axial symmetry we find the normal
deformed minimum of the ground state at a deformation $\beta\sim
0.28$, a maximum at $\beta\sim 0.52$ and a superdeformed minimum at
$\beta \sim 0.96$. We observe that the fission path (the part of blue
dashed line between normal and superdeformed minima) bypasses the
axial barrier between the normal and superdeformed  minima. The
barrier height is determined by the maximum of the  energy along this
fission path.

   The deformation energy curves for other even-even nuclei in
this region obtained in these calculations are shown in Fig.\
\ref{pot}. Full black lines show axially symmetric solutions, while
we show the values of the  deformation energy curves along the
triaxial fission path by red full curves. One can see that by
allowing for triaxial deformation the fission barrier heights are
reduced by $1-4$ MeV as compared with axially symmetric solutions.
This lowering depends on the proton and neutron numbers. It also
brings in average the results of the calculations in closer agreement
with experimental data shown by green solid circles in Fig.\
\ref{pot}. These circles display the height of the experimental
fission barrier at the calculated $\beta$-deformation of the saddle
point. The calculated $\gamma$-deformations of the triaxial parts of
the fission path are shown in Fig.\ \ref{pot-gamma}. On average they
are close to 10$^{\circ}$.

\begin{figure*}
\includegraphics[width=16.0cm]{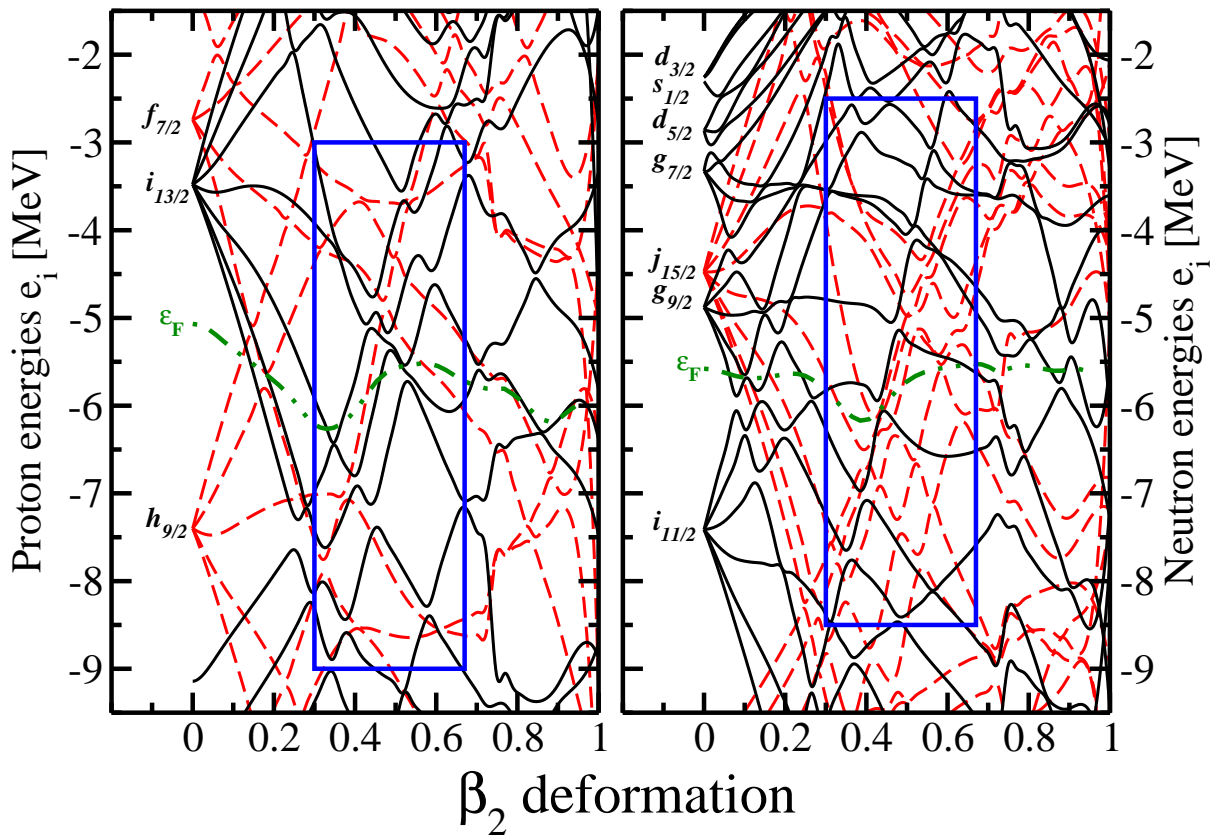}
\caption{(Color online) Proton and neutron single-particle energies
in $^{242}$Pu as a function of the quadrupole deformation $\beta_2$
for the axially symmetric solution. Solid (black) and dashed (red)
lines are used for positive and negative parity states, respectively.
Blue boxes show the regions which are displayed in more details in
Fig.\ \ref{Nilsson-ax-triax} below. Fermi energies $\varepsilon_{\rm
F}$ are shown by dot-dot-dashed (green) lines.} \label{Nilsson-diag}
\end{figure*}

\begin{figure*}
\includegraphics[width=16.0cm]{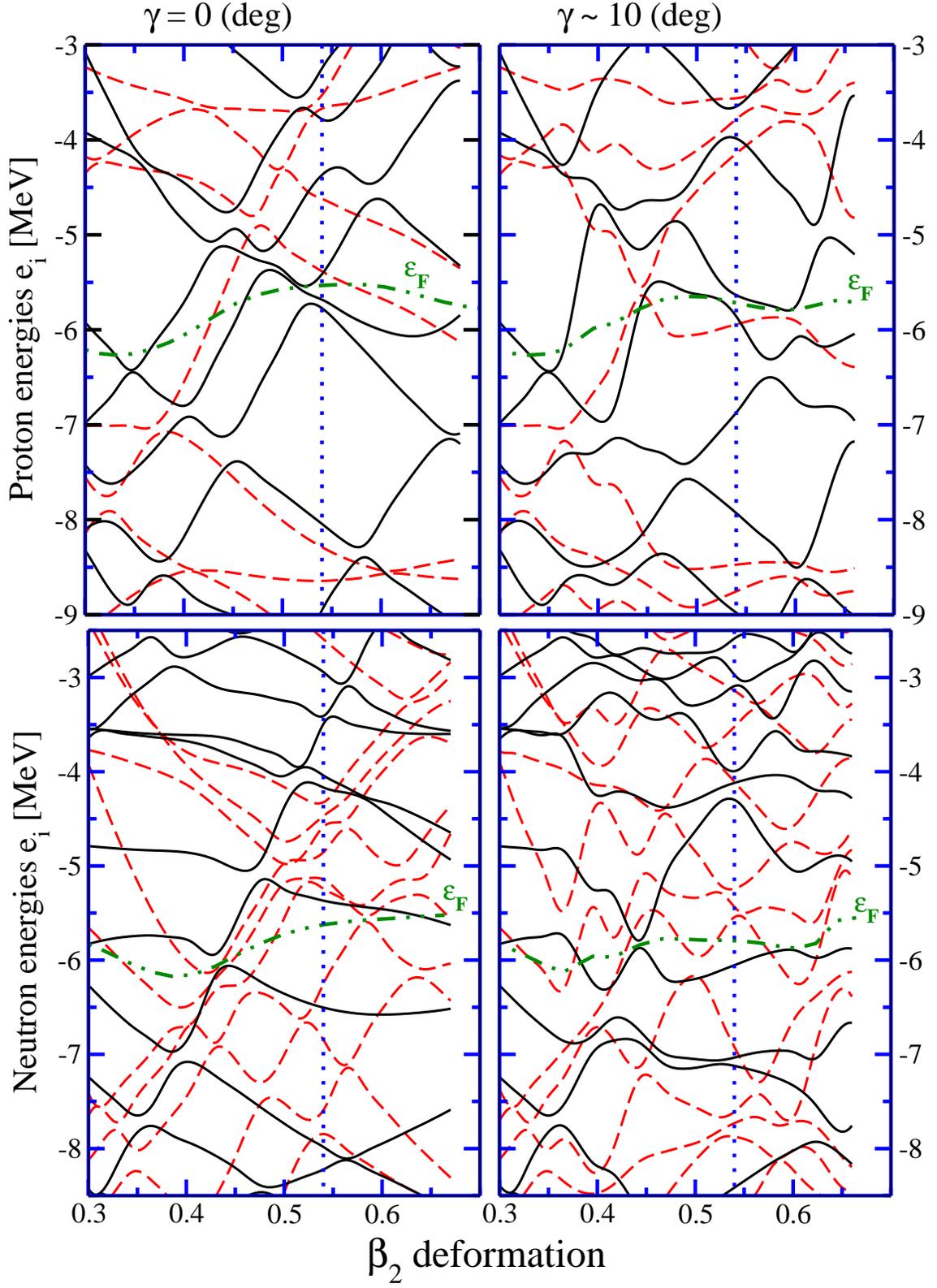}
\caption{(Color online) A comparison of proton and neutron
single-particle energies at axially symmetric (left panels) and
triaxial (right panels) solutions. The same notation for the lines as
in Fig.\ \ref{Nilsson-diag} is used. Vertical blue dotted lines are
used to show the deformation of the saddle point obtained in the
axially symmetric solution.} \label{Nilsson-ax-triax}
\end{figure*}

  The microscopic origin of the lowering of the barrier due to
triaxiality can be traced back to the changes of the level density in
the vicinity of the Fermi level induced by triaxiality. Fig.\
\ref{Nilsson-diag} shows the Nilsson diagrams for protons and
neutrons for the axially symmetric solution in $^{242}$Pu. The blue
boxes in these diagrams define the deformation and energy ranges in
which the axially symmetric and triaxial solutions are compared in
Fig.\ \ref{Nilsson-ax-triax}. The lower (upper) deformation in these
boxes corresponds to the deformation range over which the triaxial
solution (red curve in Fig.\ \ref{pot}) is lower in energy than the
corresponding axially symmetric solution (black curve in Fig.\
\ref{pot}). The lower and upper energy values in these boxes are
defined approximately as $\varepsilon_{\rm F} \pm 3$ MeV.

  Proton and neutron single-particle  energies within these deformation
and energy ranges are shown for axially symmetric and triaxial
solutions in Fig.\ \ref{Nilsson-ax-triax}. One can see that the
single-particle level density at the Fermi level is lower for
triaxial solutions than for axially symmetric solutions. This is
especially clear at the deformation corresponding to the saddle point
of the axially symmetric solution (indicated by vertical dotted blue
lines in Fig.\ \ref{Nilsson-ax-triax}) which correspond to a maximal
level density and maximal pairing correlations. A lowering of the
level density at the Fermi surface leads to a more negative shell
correction energy (as compared with axially symmetric solution), and,
as a consequence, to a lower fission barrier. This is in agreement
with the analysis of Ref.\ \cite{SK.06} which also attributes the
lowering of the inner fission barrier due to triaxiality to
microscopic (shell correction) part of the macroscopic+microscopic
model. A similar mechanism is responsible for the lowering of the
asymmetric saddle with respect to symmetric saddle at outer fission
barrier (see Sect.\ VI in Ref.\ \cite{MSI.09}).

\begin{figure}
\includegraphics[width=8.0cm]{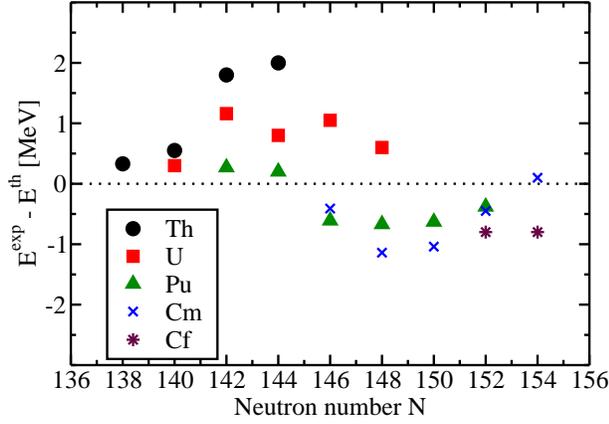}
\caption{(Color online) The difference between experimental and calculated
heights of inner fission barriers as a function of neutron number $N$.}
\label{Th-vs-exp-N}
\end{figure}

\begin{figure}
\includegraphics[width=8.0cm]{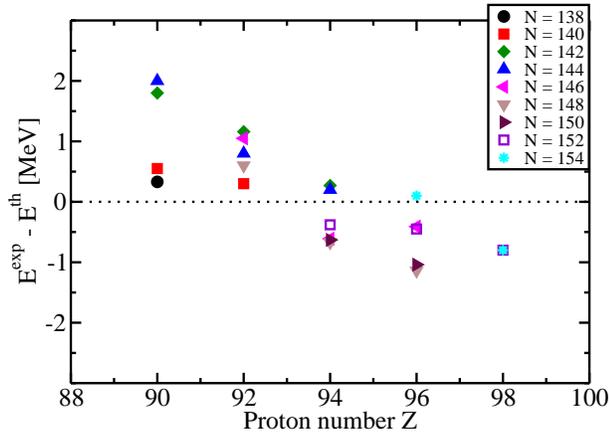}
\caption{(Color online) The same as in Fig.\ \ref{Th-vs-exp-N} but as a
function of proton number $Z$.}
\label{Th-vs-exp-Z}
\end{figure}

  Figures \ref{Th-vs-exp-N} and \ref{Th-vs-exp-Z} show the differences between
calculated and experimental heights of inner fission barriers. The average
deviation between theory and experiment is 0.76 MeV. This is comparable with
the results obtained in the macroscopic+microscopic method (see Sec. IVC
and Fig. 11 in Ref.\ \cite{DPB.07} and Sec. VII A in Ref.\ \cite{MSI.09})
which describe experimental fission barriers with an average error of around
1 MeV.

  It is necessary, however, to say that neither proton nor neutron particle
number dependences of fission barrier height are completely
reproduced in these calculations. This is clearly seen in Figs.\
\ref{Th-vs-exp-N} and \ref{Th-vs-exp-Z}. However, the same problem
exists also in macroscopic+microscopic calculations (see Fig.\  11 in
Ref.\ \cite{DPB.07} and Figs.\ 23-32 in Ref.\ \cite{MSI.09}). There
are very few energy density functional calculations of the fission
barriers with triaxiality included, and neither of them confronts in
a systematic way experimental data in actinides. However, limited
results in the Skyrme EDF presented in Ref.\ \cite{BQS.04} show
similar unresolved particle number dependences for the inner fission
barrier heights.

\section{Results for the parameter sets DD-ME2 and
DD-PC1} \label{Sec-DD}

\begin{figure}
\includegraphics[width=8.0cm]{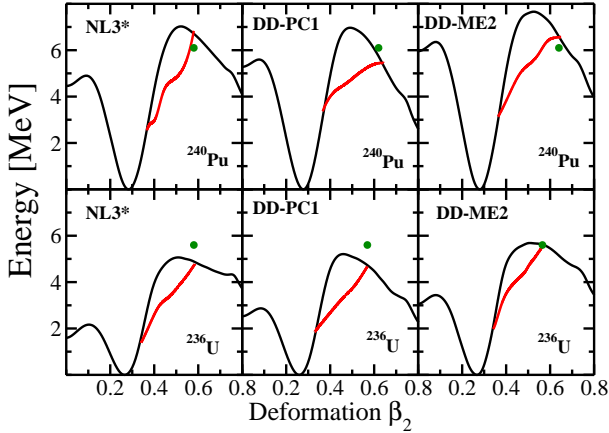}
\caption{(Color online) The same as in Fig.\ \ref{pot} but for
fission barriers in $^{240}$Pu and $^{236}$U obtained in the
calculations with the NL3*, DD-ME2, and DD-PC1 parameterizations of
the RMF Lagrangian.} \label{Diff-par}
\end{figure}

  In order to investigate to what extent our results depend on the
density functional under investigation we performed also an analysis
of the fission barriers of the two nuclei $^{240}$Pu and $^{236}$U
using the parameter sets DD-ME2 \cite{DD-ME2} and DD-PC1
\cite{NVR.08}. The first is a representative of the class of the RMF
models \cite{TW.99,DD-ME2} where the nucleus is described as a system
of Dirac nucleons interacting via the exchange of mesons with finite
masses leading to interactions of finite range. An explicit density
dependence for the meson-nucleon vertices is used. The DD-PC1
parameterization belongs to the class of the RMF models in which the
finite-range meson exchange is replaced by zero-range interactions
with density dependent coupling constants and derivative terms
\cite{NHM.92,BMM.02,NVR.08}
  .

  In Fig.\ \ref{Diff-par} we compare the deformation energy curves of $^{240}$Pu and
$^{236}$U obtained in the calculations with the three parameter sets NL3*, DD-PC1 and
DD-ME2 of the RMF Lagrangian. Although there are some differences between the
deformation energy curves obtained in the calculations with different parameterizations,
in general, they show the same features. In addition, calculated
fission barrier heights
reasonably agree with experimental data. More systematic
investigations of fission barriers with the DD-ME2 and DD-PC1
parameterizations of the RMF Lagrangian are in progress and
their results will be presented in a forthcoming manuscript.

\section{Conclusions}
\label{Sec-final}

  We presented here the first systematic investigation of triaxial
fission barriers in the actinide region within covariant density
functional theory. The calculations have been carried out with
the parameter set NL3* and they have been compared in specific cases
also with the results of parameter sets DD-ME2 and
DD-PC1. Pairing correlations are taken into account in the BCS
approximation using seniority zero forces adjusted to empirical
values of the gap parameters.  It is found that with only one
exception ($^{234}$Th) in all the nuclei under investigation the
height of the inner fission barrier is reduced by allowing for
triaxial deformations by $1-4$ MeV. The fission path avoids a
maximum of the axially symmetric potential energy surface between
the first and the second minimum by going through a valley
in the $(\beta, \gamma)$ plane with a triaxial deformation
$\gamma\approx 10^\circ$.  A systematic comparison of our results with
experimentally determined fission barriers in this region shows
reasonable agreement with data comparable with the best
macroscopic+microscopic calculations.

\section{Acknowledgements}

  This work has been supported by the U.S. Department of Energy
under the grant DE-FG02-07ER41459 and by the DFG cluster of
excellence \textquotedblleft Origin and Structure of the Universe
\textquotedblright\ (www.universe-cluster.de).

\end{document}